\documentclass[conference,a4paper]{IEEEtran}

\usepackage[innermargin=0.545in,outermargin=0.545in,top=0.6in,bottom=.7in]{geometry}
\usepackage{amsmath, amssymb,amsthm,mathtools}
\usepackage{pifont}%
\newcommand{\cmark}{\ding{51}}%
\newcommand{\xmark}{\ding{55}}%

\usepackage{arydshln}
\newtheorem{theorem}{Theorem}
\newtheorem{definition}{Definition}
\newtheorem{remark}{Remark}
\newtheorem{lemma}{Lemma}
\newtheorem{corollary}{Corollary}

\newtheorem{example}{Example}
\newtheorem{conjecture}{Conjecture}
\newtheorem{proposition}{Proposition}

\newcommand{\sfa}{\mathsf{a}}
\newcommand{\sfb}{\mathsf{b}}
\newcommand{\sfc}{\mathsf{c}}
\newcommand{\sfd}{\mathsf{d}}
\newcommand{\sfe}{\mathsf{e}}
\newcommand{\sff}{\mathsf{f}}
\newcommand{\sfg}{\mathsf{g}}

\newcommand{\bg}{\mathbf{g}}
\newcommand{\bc}{\mathbf{c}}

\newcommand{\bG}{\mathbf{G}}
\newcommand{\bH}{\mathbf{H}}
\newcommand{\bA}{\mathbf{A}}

\newcommand{\bE}{\mathbf{E}}

\newcommand{\cC}{\mathcal{C}}
\newcommand{\cE}{\mathcal{E}}
\newcommand{\cI}{\mathcal{I}}
\newcommand{\cU}{\mathcal{U}}
\newcommand{\cV}{\mathcal{V}}

\newcommand{\sTP}{\mathsf{TP}}

\newcommand{\bbE}{\mathbb{E}}
\newcommand{\bbC}{\mathbb{C}}
\newcommand{\bbCMR}{\mathbb{C}^{\mathsf{MR}}}
\newcommand{\bbI}{\mathbb{I}}
\newcommand{\bbIcol}{\mathbb{I}_{\mathsf{col}}}
\newcommand{\bbIrow}{\mathbb{I}_{\mathsf{row}}}
\newcommand{\cJ}{\mathcal{J}}

\newcommand{\cCin}{\mathcal{C}_{\mathsf{in}}}
\newcommand{\cCout}{\mathcal{C}_{\mathsf{out}}}
\newcommand{\Gin}{\bG_{\mathsf{in}}}
\newcommand{\Gout}{\bG_{\mathsf{out}}}
\newcommand{\kin}{k_{\mathsf{in}}}
\newcommand{\kout}{k_{\mathsf{out}}}

\newcommand{\nout}{n_{\mathsf{out}}}
\newcommand{\Grow}{\bG_{\mathsf{row}}}
\newcommand{\Gcol}{\bG_{\mathsf{col}}}
\newcommand{\Hrow}{\bH_{\mathsf{row}}}
\newcommand{\Hcol}{\bH_{\mathsf{col}}}
\newcommand{\Hlocal}{\bH_{\mathsf{local}}}
\newcommand{\Hglobal}{\bH_{\mathsf{global}}}
\newcommand{\cCrow}{\cC_{\mathsf{row}}}
\newcommand{\cCcol}{\cC_{\mathsf{col}}}
\newcommand{\Gab}{\mathsf{Gab}}

\newcommand{\prow}{p^{\mathsf{row}}}
\newcommand{\pcol}{p^{\mathsf{col}}}

\newcommand{\F}{\mathbb{F}}

\DeclareMathOperator{\diag}{diag}

\DeclareMathOperator{\rank}{rank}
\DeclareMathOperator{\short}{short}
\DeclareMathOperator{\punct}{punct}

\newcommand{\myspan}[1]{\left\langle #1 \right\rangle}

\newenvironment{smallpmatrix}{\left(\begin{smallmatrix}}{\end{smallmatrix}\right)}

\usepackage[dvipsnames]{xcolor}
\newif\ifcomment
\commentfalse

\title{Correctable Erasure Patterns in Product Topologies}

\author{%
  \IEEEauthorblockN{Lukas Holzbaur\IEEEauthorrefmark{1},
                    Sven Puchinger\IEEEauthorrefmark{2},
                    Eitan Yaakobi\IEEEauthorrefmark{3},
                    and Antonia Wachter-Zeh\IEEEauthorrefmark{1}}
  \IEEEauthorblockA{\IEEEauthorrefmark{1}%
                    Technical University of Munich,
                    \{lukas.holzbaur, antonia.wachter-zeh\}@tum.de}
  \IEEEauthorblockA{\IEEEauthorrefmark{2}%
                    Technical University of Denmark,
                    svepu@dtu.dk}
  \IEEEauthorblockA{\IEEEauthorrefmark{3}%
                    Technion --- Israel Institute of Technology,
                    yaakobi@cs.technion.ac.il}\vspace{-2ex}
  \thanks{The work of L. Holzbaur and A. Wachter-Zeh was supported by the German Research Foundation (Deutsche Forschungsgemeinschaft, DFG) under Grant No. WA3907/1-1. S. Puchinger has received funding from the European Union’s Horizon 2020 research and innovation programme under the Marie Skłodowska-Curie grant agreement no.~713683. The work of E. Yaakobi was partially supported by the ISF grant 1817/18 and by the Technion Hiroshi Fujiwara cyber security research center and the Israel cyber directorate.}
}

\IEEEoverridecommandlockouts
\begin{document}

\maketitle

\begin{abstract}
Locality enables storage systems to recover failed nodes from small subsets of surviving nodes. The setting where nodes are partitioned into subsets, each allowing for local recovery, is well understood. In this work we consider a generalization introduced by Gopalan et al., where, viewing the codewords as arrays, constraints are imposed on the columns and rows in addition to some global constraints. Specifically, we present a generic method of adding such global parity-checks and derive new results on the set of correctable erasure patterns. Finally, we relate the set of correctable erasure patterns in the considered topology to those correctable in tensor-product codes.
\end{abstract}

\section{Introduction}

Distributed data storage systems consist of an ever growing number of nodes/servers. To protect against data loss, the data is encoded with error/erasure correcting codes, most commonly, maximum distance separable (MDS) codes. While these offer the optimal trade-off between probability of data loss and storage overhead, the recovery of a failed node must involve a large number of nodes. This undesirable overhead lead to the introduction of \emph{locally recoverable codes} (LRCs) \cite{gopalan2012locality}, which allow for the recovery of a small number of failed nodes by contacting only a limited number of surviving nodes. In coding theoretic terms, LRCs can recover any small number of erasures from small subsets of surviving positions. For a fixed storage overhead, this improvement comes at the cost of a reduced minimum distance \cite{gopalan2012locality,rawat2013optimal}, i.e., number of failures that is guaranteed to be correctable. However, codes of optimal minimum distance are not necessarily also optimal in terms of mean time to data loss, as larger numbers of failed nodes might also be correctable. Codes that are able to recover all such combinations of failed nodes are coined \emph{maximally recoverable} (MR) \cite{chen2007maximally}. Such MR LRCs (also referred to as PMDS codes) have received considerable attention in recent years. Constructions for different parameters regimes are given in \cite{blaum2016construction,blaum2013partial, gopalan2014explicit,hu2016new,gopalan2017maximally,balaji2015partial} and the set of correctable erasure patterns has been characterized \cite{blaum2013partial,gopalan2014explicit}.

While LRCs represent an important notion of locality, some real world systems require more complex locality constraints. For example, Facebook's f4 storage system \cite{muralidhar2014f4} employs MDS codes of length $14$ tolerating $4$ failed nodes within each data center, plus an additional parity check across $3$ data centers. A theoretical framework for such restraints, referred to as the \emph{topology} of the system, was established in \cite{gopalan2014explicit}. A topology is a restriction on the support of the parity-check matrix of a code, i.e., it specifies a subset of positions in this matrix that must be zero. One case of particular interest is given by grid-like/product topologies, which, viewed as an array, impose constraints on each row and column, plus an additional number of global constraints. Even in this simplification of the general setting, it is a highly non-trivial problem to determine the erasure patterns that are correctable, i.e., the patterns that an MR code must be able to correct. The seminal works of \cite{gopalan2014explicit,gopalan2017maximally} initiated studies of the classification of these patterns for some restricted cases (see Table~\ref{tab:topologies}) \cite{shivakrishna2018maximally} and bounds on the required field size \cite{kong2019new,kane2019independence}. In particular, \cite{gopalan2017maximally} established a necessary condition for an erasure pattern to be correctable, which is then shown to also be sufficient for the case of one column constraint and no global constraints. Further, the sufficiency of this condition is conjectured to hold for the case of more column constraints.

The main results of this work are twofold. First, we establish a generic method of adding global constraints to MR codes, which allows us to establish the set of correctable erasure patterns as a function of the patterns which are correctable  without the global constraints. Second, we show that the necessary condition for correctability given in \cite{gopalan2017maximally} is \emph{not} sufficient in general. Finally, we connect MR codes for product topologies to tensor-product (TP) codes \cite{wolf1965codes}, which are another class of codes with application to storage.

\begin{table*}
  \centering
  \caption{Properties of different topologies. Lines 1-4 collect the known results. Lines 5/6 summarizes the results on global redundancy of Section~\ref{sec:globalRedundancy}, where the $\sim$ symbol implies that a code for the respective topology can be constructed given a code for the same topology with $h=0$. Lines 7/8 give the results on the characterization of correctable erasure patterns presented in Section~\ref{sec:negativeResults}.   %
  }\label{tab:topologies}
  \begin{tabular}{c|ccc|cc}
    Topology & Correctable Patterns &  Fits conj. & Reference & Construction & Reference\\ \hline
      $T_{m \times n}(0,b,h)$  & regular + any $h$ & - & \cite{blaum2013partial,gopalan2014explicit} & \cmark & \cite{rawat2013optimal,calis2016general} \\
      $T_{m \times n}(1,1,0)$  & regular & \cmark & \cite{gopalan2017maximally} & \cmark & \cite{gopalan2017maximally}  \\
      $T_{m \times n}(1,b,0)$  & regular & \cmark & \cite{gopalan2017maximally} & - & - \\
    $T_{m \times n}(1,1,h)$  & regular + any $h$ & - & \cite{gopalan2014explicit}, \cite{gopalan2017maximally} + Thm.~\ref{thm:constructionCorrectable}  & \cmark & \cite{gopalan2014explicit}, \cite{gopalan2017maximally} + Thm.~\ref{thm:constructionCorrectable}  \\
    \hdashline
        $T_{m \times n}(1,b,h)$  & regular + any $h$ & - & \cite{gopalan2017maximally} + Thm.~\ref{thm:constructionCorrectable} & $\sim$ & Thm.~\ref{thm:constructionCorrectable} + any $\cC \in \bbCMR_{m\times n}(1,b,0)$ \\
      $T_{m \times n}(a,b,h)$  & $\bbE_{m\times n}(a,b,0)$ + any $h$ & - & Thm.~\ref{thm:correctableGlobal} & $\sim$ & Thm.~\ref{thm:constructionCorrectable} + any $\cC \in \bbCMR_{m\times n}(a,b,0)$ \\
    $T_{5 \times 5}(2,2,0)$  & See Rem.~\ref{rem:permutations} & \xmark & Lem.~\ref{lem:incorrectablePattern} & - & - \\
    $T_{(\geq a+3) \times (\geq b+3)}(\geq 2,\geq 2,0)$  & unknown & \xmark & Thm.~\ref{thm:largerPatterns} & - & - \\
  \end{tabular}
  \vspace*{-15pt}
\end{table*}

\section{Preliminaries}

\subsection{Notation}
We denote the set of integers $[a,b] = \{i \ | \ a\leq i \leq b\}$ and write $[1,b] = [b]$.
For an $a \times b$ matrix $\bA$ and a set $\cI \subset [b]$ we denote by $\bA|_{\cI}$ the restriction of $\bA$ to the columns indexed by $\cI$. We write $\mathbf{0}_{b}$ to denote the all-zero vector of length $b$.
Let $\F_q$ denote the finite field of size $q$.
We denote a code $\cC$ of length $n$ and dimension $\dim(\cC) = k$ over $\F_q$ by $[n,k]_q$. If the field size $q$ is not of interest we write $\F$ and $[n,k]$. A set $\cI \subseteq [n]$ with $|\cI|=k$ is said to be an \emph{information set} of $\cC$ if $\dim(\cC|_\cI) = k$.
For sets $\cU \subseteq [m]$ and $\cV \subseteq [n]$ we denote the Cartesian product $\cU \times \cV = \{ (U,V) \ | \ U \in \cU, V \in \cV \}$.
For two codes  $\cC_1 \subseteq \F^{m}$ and $\cC_2 \subseteq \F^{n}$ we write
$\cC_1\otimes \cC_2 = \myspan{\bG_1 \otimes \bG_2}$ , %
where $\bG_1, \bG_2$ denote some generator matrix of $\cC_1,\cC_2$, respectively, $\otimes$ is the Kronecker product, and $\myspan{\bG}$ is the row span of the matrix $\bG$. Note that the resulting code is independent of the specific choice of the generator matrices. We have $\cC_1 \otimes \cC_2 \subseteq \F^{mn}$ but we frequently use the equivalent interpretation as a subset of~$\F^{m\times n}$. %

\subsection{Codes for Grid-Like Topologies}

In this work we are concerned with codes for grid-like topologies $T_{m \times n}(a,b,h)$, i.e., non-rigorously, product codes with $h$ additional parity constraints that are allowed to depend on all codeword positions. We adapt the notation of \cite{gopalan2017maximally}.

\begin{definition}[Code for grid-like topology {\cite[Definition~2.1]{gopalan2017maximally}}]\label{def:codesForGridLike}
  Let $\cCcol$ be an $[m,\geq m-a]$ code and $\cCrow$ be an $[n,\geq n-b]$ code. We define a code for the topology $T_{m \times n}(a,b,h)$ to be any code with parity-check matrix
  \begin{align*}
    \bH =
    \begin{pmatrix}
      \Hlocal \\ \Hglobal
    \end{pmatrix} \ ,
  \end{align*}
  where $\Hlocal$ is a parity-check matrix of the code $\cCcol \otimes \cCrow$ and $\Hglobal$ is an arbitrary $h\times mn$ matrix.

  We denote the set of all such codes by $\bbC_{m\times n}(a,b,h)$.
\end{definition}

\begin{definition}[Correctable erasure pattern {\cite[Definition~2.2]{gopalan2017maximally}}]\label{def:correctableErasurePatterns}
  Let $\cE\subseteq [m] \times [n]$ denote a set of erased positions. We say the \emph{erasure pattern} $\cE$ is \emph{correctable in the topology} $T_{m \times n}(a,b,h)$ if and only if there exists a code in $\bbC_{m\times n}(a,b,h)$ that can correct this erasure pattern.

  We denote the set of erasure patterns which are correctable in $T_{m \times n}(a,b,h)$ by $\bbE_{m\times n}(a,b,h)$ and by $\bbE^{\max}_{m\times n}(a,b,h)$ those that are not a proper subset of any other correctable pattern.
\end{definition}

\begin{definition}[Maximally recoverable code~{\cite[Definition~2.3]{gopalan2017maximally}}]\label{def:MRcode}
  We say a code $\cC \in \bbC_{m\times n}(a,b,h)$ is \emph{maximally recoverable} (MR) if it corrects every erasure pattern in $\bbE_{m\times n}(a,b,h)$.

  We denote the set of codes that are MR for a topology $T_{m \times n}(a,b,h)$ by $\bbC_{m\times n}^{\mathsf{MR}}(a,b,h)$.
\end{definition}

Observe that, any code that can correct an erasure pattern $\cE$ can also correct any erasure pattern $\cE'\subset \cE$. A maximally recoverable code is equivalently defined as being able to correct any pattern in $\bbE^{\max}_{m\times n}(a,b,h)$ instead of $\bbE_{m\times n}(a,b,h)$. By the same argument, the set of correctable erasure patterns $\bbE_{m\times n}(a,b,h)$ is uniquely defined by $\bbE^{\max}_{m\times n}(a,b,h)$.

We summarize several important properties of codes for grid-like topologies in terms of our notation. Throughout this work we let $m,n,a,b,h$ be non-negative integers which satisfy $m>a$ and $n>b$. Further, to exclude trivial cases where the dimension of the code $\cC$ is smaller than the dimension of the codes $\cCcol$ and $\cCrow$, we assume that $h \leq (m-a)(n-b)-\max\{m-a,n-b\}$ for the remainder of this work.

\begin{proposition}[Properties of codes for grid-like topologies~{\cite[Proposition~2.1]{gopalan2017maximally}}]\label{prop:propertiesGridLikeMR}
  For any $\cC \in \bbCMR_{m\times n}(a,b,h)$ it holds that
  \begin{itemize}
  \item the dimensions of $\cC, \cCcol$, and $\cCrow$ are
    \begin{align*}
      \dim(\cC) &= (m-a)(n-b)-h \\
      \dim(\cCcol) = m&-a \quad \text{and} \quad \dim(\cCrow) = n-b \ ,
    \end{align*}
  \item the codes $\cCcol$ and $\cCrow$ are MDS.
  \end{itemize}
\end{proposition}

\begin{definition}[{\cite[Definition~3.1]{gopalan2017maximally}}]\label{def:regularPattern}
  Consider the topology $T_{m\times n}(a,b,0)$ and an erasure pattern $\cE \subset [m] \times [n]$. We say that $\cE$ is regular\footnote{Note that the original definition does not include the restriction $u\geq a$ and $v \geq b$. However, it is easy to check that this restriction is indeed necessary.} if for all $\cU\subseteq [m], |\cU|=u\geq a$, and $\cV\subseteq [n], |\cV|=v\geq b$, we have
  \begin{align*}
    |\cE\cap (\cU\times \cV)| \leq va + ub - ab \ .
  \end{align*}
\end{definition}

\begin{remark}
  Intuitively, the restriction of the erasure pattern to $\cU\times \cV$ can be interpreted as the shortening of the respective code in the positions outside of this grid. As we will show in Section~\ref{sec:implications}, this shortened code needs to be able to decode the remaining erasures for the pattern to be correctable. Definition~\ref{def:regularPattern} is a necessary (in general insufficient, see Section~\ref{sec:counterExample}) condition for this to be possible, namely, that the number of erasures remaining in the shortened code does not exceed its redundancy.
\end{remark}

It was shown in \cite{gopalan2017maximally} that all erasure patterns that are \emph{not} regular are \emph{not} correctable in the topology $T_{m\times n}(a,b,0)$. On the other hand, for some cases (see Table~\ref{tab:topologies}) it is known that \emph{all} regular patterns are correctable, which led to the following conjecture.

\begin{conjecture}[{\cite[Conjecture~3.1]{gopalan2017maximally}}]\label{con:regularCorrectable}
  An erasure pattern $\cE$ is correctable for the topology $T_{m \times n}(a,b,0)$ if and only if it is regular.
\end{conjecture}
In Section~\ref{sec:counterExample} we disprove this conjecture.

\section{Global Redundancy $h>0$} \label{sec:globalRedundancy}

For some topologies with $h=0$ the set of correctable erasure patterns is fully characterized (as the regular patterns) and, for some cases, explicit constructions are known. For $h>0$ only the special cases of $a=0 ,b,h\geq 0$ and $a=b=h=1$ are known (see Table~\ref{tab:topologies}). In this section, we characterize the set of correctable erasure patterns $\bbE^{\max}_{m\times n}(a,b,h)$ as a function of $\bbE^{\max}_{m\times n}(a,b,0)$, for any $m,n,a,b$, and $h$. A similar result for the extension of codes defined by a binary parity-check matrix, i.e., where $\Hlocal$ is in $\F_2$, has been derived in \cite[Section~V.A]{gopalan2014explicit}.

\subsection{Construction}

We generalize the Gabidulin-based code construction~\cite{rawat2013optimal} for $a=0$. These codes have been shown to be MR for the topology $T_{m \times n}(0,b,h)$ in~\cite{calis2016general}. We show that by the same two-stage encoding procedure we can ``add global redundancy'' to codes for any grid-like topology. In other words, we give a general construction that, given a code $\cCout \in \bbCMR_{m\times n}(a,b,0)$, returns a code $\cC \in \bbCMR_{m\times n}(a,b,h)$.

\begin{definition}[Gabidulin codes]\label{def:GabidulinCodes}
Let $\bg = (g_1,\ldots, g_n) \in \F_{q^s}^{n}$ be such that the $g_i$ are linearly independent over $\F_q$.
The $[n,k]_{q^s}$ Gabidulin code $\Gab(n,k,\bg)$ is defined as
\begin{align*}
  \Gab(n,k,\bg) = \myspan{\bG}
\end{align*}
with
\begin{align*}
\bG =
\begin{pmatrix}
    g_{1} & g_{2} & \hdots & g_{n}\\
    g_{1}^{q^1} & g_{2}^{q^1} & \hdots & g_{n}^{q^1}\\
    \vphantom{\int\limits^x}\smash{\vdots} & \vphantom{\int\limits^x}\smash{\vdots} & \ddots & \vphantom{\int\limits^x}\smash{\vdots} \\
    g_{1}^{q^{k-1}} & g_{2}^{q^{k-1}} & \hdots & g_{n}^{q^{k-1}}
\end{pmatrix} \ .
\end{align*}
\end{definition}

Observe that Gabidulin codes exist for any $n\leq s$.
Gabidulin codes are designed for the rank metric, but it is also well-known that they are MDS in the Hamming metric.
We recall a well-known property of Gabidulin codes.

\begin{lemma}[{\cite[Lemma~3]{berger2003isometries}}]\label{lem:GabidulinCodeLocatorTransformation}
  Let $\bG\in \F_{q^s}^{k\times n}$ be a generator matrix of an $[n,k,d]$ Gabidulin code $\Gab(n,k,\bg)$. Then, for any full-rank matrix $\bA \in \F_q^{n\times n}$, the code
    $\cC' = \left\langle \bG \cdot \bA \right\rangle$
  is an $[n,k,d]$ Gabidulin code $\Gab(n,k,\bg')$ with $\bg' =  \bg \cdot \bA$.
\end{lemma}

\begin{lemma}\label{lem:gabidulinConstruction}
  Let $\bbI_{\mathsf{out}}$ be the set of information sets of $\cCout[\nout,\kout]_q$ and $\cCin$ be a $\Gab(\kout,\kin,\bg)_{q^s}$ Gabidulin code. Then the code
    $\myspan{\Gin \cdot \Gout} |_{\cI}$
  is a $\Gab(\kout,\kin,\bg')_{q^s}$ Gabidulin code for any $\cI \in \bbI_{\mathsf{out}}$.
\end{lemma}
\begin{IEEEproof}
  By definition of an information set, the matrix $\Gout|_{\cI}$ is of full rank. The statement follows from observing that $\myspan{\Gin \cdot \Gout} |_{\cI} = \myspan{\Gin \cdot (\Gout |_{\cI})}$ and applying Lemma~\ref{lem:GabidulinCodeLocatorTransformation}.
\end{IEEEproof}

\begin{theorem}\label{thm:constructionCorrectable}
  Let $\cCout \in \bbCMR_{m\times n}(a,b,0)$ be an $[\nout = mn,\kout =(m-a)(n-b)]$ code and
  \begin{equation*}
      \cCin = \Gab\big((m-a)(n-b),(m-a)(n-b)-h,\bg \big)_{q^s} \ ,
  \end{equation*}
  with $s\geq (m-a)(n-b)$. Then the code $\myspan{\Gin \cdot \Gout}$ corrects all erasure patterns in
  \begin{align*}
       \{ \cE' \cup \cI &\ \mid \ \cE' \in \bbE^{\max}_{m\times n}(a,b,0),\cI \subset ([m]\times [n] \setminus \cE'), |\cI| = h\} \ .
  \end{align*}
\end{theorem}
\begin{IEEEproof}
Let $\cE$ be an erasure pattern of the set above.
Then, there is an erasure pattern $\cE'$ of the set $\bbE^{\max}_{m\times n}(a,b,0)$ with $\cE = \cE' \cup \cI$.
The complement of $\cE'$ is, by definition, an information set of the outer code $\cCout$.
By restricting the overall code to this information set, we thus obtain a Gabidulin code with parameters $[(m-a)(n-b),(m-a)(n-b)-h]$ by Lemma~\ref{lem:gabidulinConstruction}.
Note that there are exactly $h$ remaining erasures (given by $\cI$) in the remaining positions.
Since any Gabidulin code is MDS, the restricted code can correct exactly $h$ erasures, which concludes the proof.
\end{IEEEproof}

\begin{example}
  Let $\cCout[m n, m (n-b)]$ be the code spanned by
  \begin{align*}
    \diag(\underbrace{\bG,\bG,\ldots,\bG}_{m \text{ times}}) \ ,
  \end{align*}
where $\bG$ is the generator matrix of an arbitrary $[n,n-b]_q$ MDS code. Observe that $\cCout \in \bbCMR_{m\times n}(0,b,0)$ and the set of its information sets is given by
  \begin{align*}
    \bbI_{\mathsf{out}} = \{\text{From each block pick arbitrary } n-b \text{ positions}\} \ .
  \end{align*}
  Choose $\cCin$ to be an $[m (n-b),m (n-b)-h]_{q^{m (n-b)}}$ Gabidulin code to obtain the PMDS code construction of \cite{rawat2013optimal}. Observe that $\bbI_{\mathsf{out}}$ are exactly the subsets of positions that must give an MDS code in a PMDS code.
\end{example}

\subsection{Correctable Erasure Patterns}
Let us characterize the patterns correctable in $T_{m\times n}(a,b,h)$ given the set of patterns correctable in $T_{m\times n}(a,b,0)$.

\begin{theorem}\label{thm:correctableGlobal}
We have
   \begin{align*}
      \bbE^{\max}_{m\times n}(a,b,h) = \{ \cE' \cup \cI &\ \mid \ \cE' \in \bbE^{\max}_{m\times n}(a,b,0),\\
      &\cI \subset (m\times n \setminus \cE'), |\cI| = h\} \ ,
    \end{align*}
    i.e., any $\cE\in \bbE^{\max}_{m\times n}(a,b,h)$ can be obtained by adding $h$ erasures to some $\cE' \in \bbE^{\max}_{m\times n}(a,b,0)$.
\end{theorem}

\begin{IEEEproof}
``$\supseteq$'' follows by the construction of Theorem~\ref{thm:constructionCorrectable}.

The other direction is implied by the following argument.
Let $\cE \in \bbE^{\max}_{m\times n}(a,b,h)$. Denote by $\bG$ the generator matrix of a code $\cC \in \bbC_{m\times n}(a,b,h)$ that corrects $\cE$ and let $\Hlocal,\Hglobal$ be as in Definition~\ref{def:codesForGridLike}. Denote by $\cC' \in \bbC_{m\times n}(a,b,0)$ the code obtained by setting $\Hglobal = \mathbf{0}$. Then there exists a generator matrix of $\cC'$ of the form
\begin{align*}
  \bG' =
  \begin{pmatrix}
    \bG \\
    \bG_{\mathsf{global}}
  \end{pmatrix} \ ,
\end{align*}
for some $\bG_{\mathsf{global}} \in \F^{h\times mn}$. Trivially, we have
\begin{align*}
\rank(\bG'|_{[mn]\setminus \cE}) \geq \rank(\bG|_{[mn]\setminus \cE}) = \dim(\cC) \ ,
\end{align*}
where the last equality holds because $\cE$ is correctable in $\cC$ by definition. By basic linear algebra arguments there exists a subset $\cI \subset \cE$ with $|\cI|=\dim(\cC)-\dim(\cC') \leq h$ such that
\begin{align*}
\rank(\bG'|_{([mn]\setminus \cE) \cup \cI}) = \rank(\bG') = \dim(\cC') \ .
\end{align*}
It follows that $\cE\setminus \cI$ is correctable in $T_{m \times n}(a,b,0)$. This concludes the proof.
\end{IEEEproof}

\begin{corollary}
  Denote $N=mn$.  Then, for any $\cE' \in \bbE^{\max}_{m\times n}(a,b,0)$ and any $[N,k]$ code $\cC \in \bbCMR_{m\times n}(a,b,h)$ the code $\cC|_{m\times n\setminus \cE'}$ must be an $[N-|\cE'|,k,h+1]$ MDS code.
\end{corollary}

\begin{IEEEproof}
  It follows trivially from the dimension of $\cC$ that the code $\cC|_{[n]\setminus \cE'}$ can never correct more than $h$ erasures. The existence of a code for which this restriction is an MDS code follows from Corollary~\ref{cor:existenceGlobalRed}.
\end{IEEEproof}

\begin{corollary}\label{cor:existenceGlobalRed}
  Let $\cCout[N,k]_{q} \in \bbCMR_{m\times n}(a,b,0)$. Then there exists an $[N,k-h]_{q^{k}}$ code $\cC \in \bbCMR_{m\times n}(a,b,h)$.
\end{corollary}

\section{Negative Results on Correctable Erasure Patterns}\label{sec:negativeResults}

In this section we turn to providing negative results on the set of correctable erasure patterns for given topology. First, we prove that a specific erasure pattern is never correctable in the topology $T_{5 \times 5}(2,2,0)$, thereby disproving Conjecture~\ref{con:regularCorrectable} as given in \cite{gopalan2017maximally}. Then, using this result, we provide generic methods of constructing uncorrectable, regular erasure patterns for larger topologies.

\subsection{Disproving a Conjecture on the Correctability of Regular Erasure Patterns}\label{sec:counterExample}
We begin with some general observations on the relation between low-weight codewords.
\begin{lemma}\label{lem:symbolRelation}
  Consider a linear $[5,3]$ code $\cCrow$ with generator matrix
  \begin{align*}
  \Grow &= \begin{pmatrix}
    1&0&0&\prow_{1,1} &\prow_{1,2}\\
   0&1&0&\prow_{2,1} &\prow_{2,2}\\
   0&0&1&\prow_{3,1} &\prow_{3,2}\\
 \end{pmatrix}
  \end{align*}
  and a codeword
  \begin{align}
  \begin{pmatrix}
    1 & \alpha_2 & \alpha_3 & 0 & 0
  \end{pmatrix} \in \cCrow \ . \label{eq:minWeightRow1}
  \end{align}
  Then
  \begin{align}
  \prow_{1,1} &= -(\alpha_2 \prow_{2,1} +\alpha_3 \prow_{3,1}) \label{eq:rowParities1}\\
  \prow_{1,2} &= -(\alpha_2 \prow_{2,2} +\alpha_3 \prow_{3,2}) \ . \label{eq:rowParities2}
\end{align}
\end{lemma}
\begin{IEEEproof}
  The statement follows from
\begin{align*}
  \begin{pmatrix}
    1 & \alpha_2 & \alpha_3 & 0 & 0
  \end{pmatrix} =
  \begin{pmatrix}
    1 & \alpha_2 & \alpha_3
  \end{pmatrix} \cdot
  \begin{smallpmatrix}
   1&0&0&\prow_{1,1} &\prow_{1,2}\\
   0&1&0&\prow_{2,1} &\prow_{2,2}\\
   0&0&1&\prow_{3,1} &\prow_{3,2}\\
 \end{smallpmatrix} \\
  \Rightarrow \
  \begin{pmatrix}
    1 & \alpha_2 & \alpha_3
  \end{pmatrix} \cdot
  \begin{smallpmatrix}
  \prow_{1,1} &\prow_{1,2}\\
  \prow_{2,1} &\prow_{2,2}\\
  \prow_{3,1} &\prow_{3,2}\\
  \end{smallpmatrix}
  =
  \begin{pmatrix}
    0&0
  \end{pmatrix} \ .
\end{align*}
\end{IEEEproof}

Observe that each row of $\Grow$ is a codeword of $\cCrow$, in particular,
\begin{align}
  \begin{pmatrix}
    1 & 0 & 0 & \prow_{1,1} & \prow_{1,2}
  \end{pmatrix} \in \cCrow \ .\label{eq:minWeightRow2}
\end{align}

We define the same notions in similar notation for a $[5,3]$ code $\cCcol$, i.e.,
\begin{align}
  &\begin{pmatrix}
    1 & \gamma_2 & \gamma_3 & 0 & 0
  \end{pmatrix} \in \cCcol \label{eq:minWeightCol1}\\
  &\begin{pmatrix}
    1 & 0 & 0 & \pcol_{1,1} & \pcol_{1,2}
  \end{pmatrix} \in \cCcol \label{eq:minWeightCol2}\\
  &\pcol_{1,1} = -(\gamma_2 \pcol_{2,1} +\gamma_3 \pcol_{3,1}) \label{eq:colParities1}\\
  &\pcol_{1,2} = -(\gamma_2 \pcol_{2,2} +\gamma_3 \pcol_{3,2}) \ . \label{eq:colParities2}
\end{align}

Note that if $\cCrow$ ($\cCcol$) is MDS we have $\prow_{1,1},\prow_{1,2} \neq 0$ ($\pcol_{1,1},\pcol_{1,2} \neq 0$) and the codewords given in (\ref{eq:minWeightRow1}) and (\ref{eq:minWeightRow2}) ((\ref{eq:minWeightCol1}) and (\ref{eq:minWeightCol2})) are unique. This follows from the well-known facts that all symbols in the parity part of a systematic generator matrix of an MDS code must be non-zero and that every codeword of weight $d$ of a linear code is unique up to scalar multiplication with an element of $\F^*$.

With these general relations established, we are now ready to prove that there exists a regular erasure pattern that is never correctable in the topology  $T_{5\times 5}(2,2,0)$, by constructing a non-zero codeword that is zero in all non-erased positions.
\begin{lemma}\label{lem:incorrectablePattern}
The regular pattern $\cE$ given by
\begin{align*}
  \bE &=
  \begin{smallpmatrix}
    0&\star&\star&\star&\star\\
    \star&\star&\star&0&0\\
    \star&\star&\star&0&0\\
    \star&0&0&\star&\star\\
    \star&0&0&\star&\star\\
  \end{smallpmatrix}\\
  \cE &= \{(i,j) \ | \ \bE(i,j) = \star \}
\end{align*}
is not correctable in $T_{5\times 5}(2,2,0)$.
\end{lemma}
\begin{IEEEproof}
Let $\cC \in \bbC_{m\times n}(a,b,0)$.
We prove the statement by constructing a non-zero codeword of $\cC$ that is zero in all non-erased positions.
This implies that the code $\cC$ cannot correct the pattern uniquely since there are at least two codewords that coincide on all non-erased positions---the constructed non-zero codeword and the all-zero codeword.

To construct such a codeword, we replace the $\star$-symbols in $\bE$ by elements of $\F$ (not all zero) in the following manner:
\begin{itemize}
\item[$(\sfa)$] Choose the element in position $(2,1)$ to be $\gamma_2\in \F^*$. Note that, since the code is linear, we can always normalize one single non-zero position to be an arbitrary element of $\F^*$ and w.l.o.g. we assume position $(2,1)$ to be non-zero.
\item[$(\sfb)$] Choose the second row to be the $\gamma_2$-multiple of (\ref{eq:minWeightRow1}).
\item[$(\sfc)$] Choose the second and third column to be the corresponding multiples of (\ref{eq:minWeightCol1}).
\item[$(\sfd)$] To obtain a multiple of (\ref{eq:minWeightRow1}) as the third row, set its first position to be $\gamma_3$.
\item[$(\sfe)$] Encode the first row and column with $\Grow$ and $\Gcol$, respectively.
\item[$(\sff)$] Replace the entries according to (\ref{eq:rowParities1}),(\ref{eq:rowParities2}),(\ref{eq:colParities1}), and (\ref{eq:colParities2}).
\item[$(\sfg)$] Fill in the fourth and fifth rows with the corresponding multiples of (\ref{eq:minWeightRow2}).
\end{itemize}
The individual steps are given by
\begin{align*}
  &\stackrel{(\sfa)}{\Rightarrow}
  \begin{smallpmatrix}
    0&\star&\star&\star&\star\\
    \gamma_2&\star&\star&0&0\\
    \star&\star&\star&0&0\\
    \star&0&0&\star&\star\\
    \star&0&0&\star&\star\\
  \end{smallpmatrix}
  \stackrel{(\sfb)}{\Rightarrow}
  \begin{smallpmatrix}
    0&\star&\star&\star&\star\\
    \gamma_2&\gamma_2 \alpha_2&\gamma_2\alpha_3&0&0\\
    \star&\star&\star&0&0\\
    \star&0&0&\star&\star\\
    \star&0&0&\star&\star\\
  \end{smallpmatrix} \\
  &\stackrel{(\sfc)}{\Rightarrow}
  \begin{smallpmatrix}
    0&\alpha_2&\alpha_3&\star&\star\\
    \gamma_2&\gamma_2 \alpha_2&\gamma_2\alpha_3&0&0\\
    \star&\gamma_3 \alpha_2&\gamma_3 \alpha_3&0&0\\
    \star&0&0&\star&\star\\
    \star&0&0&\star&\star\\
  \end{smallpmatrix}
  \stackrel{(\sfd)}{\Rightarrow}
  \begin{smallpmatrix}
    0&\alpha_2&\alpha_3&\star&\star\\
    \gamma_2&\gamma_2 \alpha_2&\gamma_2\alpha_3&0&0\\
    \gamma_3&\gamma_3 \alpha_2&\gamma_3 \alpha_3&0&0\\
    \star&0&0&\star&\star\\
    \star&0&0&\star&\star\\
  \end{smallpmatrix} \\
  &\stackrel{(\sfe)}{\Rightarrow}
  \begin{smallpmatrix}
    0&\alpha_2&\alpha_3&\alpha_2\prow_{2,1}+\alpha_3\prow_{3,1}&\alpha_2\prow_{2,2}+\alpha_3\prow_{3,2}\\
    \gamma_2&\gamma_2 \alpha_2&\gamma_2\alpha_3&0&0\\
    \gamma_3&\gamma_3 \alpha_2&\gamma_3 \alpha_3&0&0\\
    \gamma_2\pcol_{2,1}+\gamma_3\pcol_{3,1}&0&0&\star&\star\\
    \gamma_2\pcol_{2,2}+\gamma_3\pcol_{3,2}&0&0&\star&\star\\
  \end{smallpmatrix} \\
  &\stackrel{(\sff)}{\Rightarrow}
  \begin{smallpmatrix}
    0&\alpha_2&\alpha_3&-\prow_{1,1}&-\prow_{1,2}\\
    \gamma_2&\gamma_2 \alpha_2&\gamma_2\alpha_3&0&0\\
    \gamma_3&\gamma_3 \alpha_2&\gamma_3 \alpha_3&0&0\\
    -\pcol_{1,1}&0&0&\star&\star\\
    -\pcol_{1,2}&0&0&\star&\star\\
  \end{smallpmatrix} \\
  &\stackrel{(\sfg)}{\Rightarrow}
  \begin{smallpmatrix}
    0&\alpha_2&\alpha_3&-\prow_{1,1}&-\prow_{1,2}\\
    \gamma_2&\gamma_2 \alpha_2&\gamma_2\alpha_3&0&0\\
    \gamma_3&\gamma_3 \alpha_2&\gamma_3 \alpha_3&0&0\\
    -\pcol_{1,1}&0&0&-\pcol_{1,1}\prow_{1,1}&-\pcol_{1,1}\prow_{1,2}\\
    -\pcol_{1,2}&0&0&-\pcol_{1,2}\prow_{1,1}&-\pcol_{1,2}\prow_{1,2}\\
  \end{smallpmatrix} \ .
\end{align*}
It is easy to see that the fourth and fifth columns are also multiples of (\ref{eq:minWeightCol2}). Hence, this array contains a codeword of the row code in every row and a codeword of the column code in every column. It is therefore a valid codeword of any code for $T_{5 \times 5}(2,2,0)$. As the used properties hold for any linear code we conclude that this pattern is never correctable.

We note that if both $\cCrow$ and $\cCcol$ are MDS, the first step, i.e., choosing the element in position $(2,1)$, determines the whole matrix, as each of the subsequent steps is unique in this case.
\end{IEEEproof}

\begin{remark}\label{rem:permutations}
  The proof of Lemma~\ref{lem:incorrectablePattern} can be carried out for any column/row permutation of $\bE$, i.e., all these permutations are not correctable in $T_{5\times 5}(2,2,0)$. Further, computer search shows that these $450$ permutations are \emph{exactly} the regular patterns that are not correctable in this topology.
\end{remark}

\begin{theorem}
  Conjecture~\ref{con:regularCorrectable} (cf. \cite[Conjecture~3.1]{gopalan2017maximally}) is false.
\end{theorem}
\begin{IEEEproof}
  Follows immediately from Lemma~\ref{lem:incorrectablePattern}.
\end{IEEEproof}

\subsection{Implications of Uncorrectable Regular Patterns for Larger Topologies}\label{sec:implications}
In this section, we show that the incorrectable erasure pattern for the topology $T_{5 \times 5}(2,2,0)$ given in Lemma~\ref{lem:incorrectablePattern} has implications on the set of correctable erasure for almost all topologies. To this end, we employ arguments based on \emph{shortening} and \emph{puncturing}.
\begin{definition}\label{def:shortPunct}
  Let $\cC$ be an $[N,k]$ code and $\cI \subseteq [N]$ be a set of intgers.
  We define the \emph{shortening} operator as
\begin{align*}
  \short_\cI(\cC) = \{ \bc|_{[N]\setminus \cI} \ | \ \bc \in \cC, \bc|_\cI = \mathbf{0}_{|\cI|} \}
\end{align*}
and the \emph{puncturing} operator as
\begin{align*}
  \punct_\cI(\cC) = \{ \bc|_{[N]\setminus \cI} \ | \ \bc \in \cC \}  \ .
\end{align*}
\end{definition}
We collect some well-known/basic properties in the following proposition.

\begin{proposition}\label{prop:shortPunctProperties}
  Let $\bG$ and $\bH$ denote a generator and parity-check matrix of an $[N,k]$ code $\cC$, respectively. Then,
  \begin{enumerate}
  \item $\bH|_{[N]\setminus \cI}$ is a parity-check matrix of $\short_\cI(\cC)$ and $\bG|_{[N] \setminus \cI}$ is a generator
    matrix of $\punct_{\cI}(\cC)$. \label{item:matrices}
  \item an erasure pattern $\cE \subset [N]$ is correctable if and only if it fulfills the equivalent conditions \label{item:correctable}
    \begin{align*}
      \dim(\short_{[N]\setminus\cE}(\cC)) &= 0 \\
      \dim(\punct_{\cE}(\cC)) &= k \ .
    \end{align*}
\item  an erasure pattern $\cE \subset [N]$ is correctable only if the pattern $\cE \setminus \cI$ is correctable in the code $\short_{\cI}(\cC)$ for any $\cI \subset [N]$.\label{item:correctSubpattern}
\item an erasure pattern $\cE$ is correctable only if the pattern $\cE \setminus \cI$ is correctable in the code $\punct_\cI(\cC)$ for any $\cI \subseteq [\cE]$.\label{item:incorPuncture}
\end{enumerate}
\end{proposition}
\begin{IEEEproof}
  Properties \ref{item:matrices} and \ref{item:correctable} are well-known.

\emph{Proof of \ref{item:correctSubpattern}):} By property \ref{item:correctable}) we known that $\cE \setminus \cI$ is correctable in $\short_{\cI}(\cC)$ if and only if $\short_{[N] \setminus (\cE\setminus \cI)}(\short_{\cI}(\cC) )= 0$. Now observe that
  \begin{align*}
    \short_{[N] \setminus (\cE\setminus \cI)}(\short_{\cI}(\cC) ) &= \short_{([N] \setminus (\cE\setminus \cI)) \cup \cI}(\cC)\\
                                                                  &= \short_{([N] \setminus \cE) \cup \cI}(\cC)\\
                                                                  &= \short_{\cI \setminus ([N] \setminus \cE)}(\short_{([N] \setminus \cE) }(\cC))\ .
  \end{align*}
As shortening does not increase the code dimension we have
  \begin{align*}
    \dim(\short_{[N] \setminus (\cE\setminus \cI)}(\short_{\cI}(\cC) )) \leq \dim(\short_{[N] \setminus\cE}(\cC)) \ .
  \end{align*}
  The statement follows from observing that $\cE$ is correctable in $\cC$ if and only if $\dim(\short_{[N] \setminus\cE}(\cC))=0$.

 \emph{Proof of \ref{item:incorPuncture})} By property \ref{item:correctable}) an erasure pattern is correctable if and only if $\dim(\punct_{\cE}(\cC)) = \dim(\cC)$. The statement follows from observing that $\punct_{\cE\setminus \cI}(\punct_{\cI}(\cC))  = \punct_\cE(\cC)$.
\end{IEEEproof}

\begin{lemma}\label{lem:topologyRelation}
Let $\cC \in \bbC_{m\times n}(a,b,0)$. Denote by $\bbIcol$ and $\bbIrow$ the sets of information sets of the respective column and row code. Then for any $\cU\subseteq [m]$ such that $([m] \setminus \cU) \subseteq \cI$ for some $\cI \in \bbIcol$ and $\cV\subseteq [n]$ such that $([n] \setminus \cV) \subseteq \cI$ for some $\cI \in \bbIrow$ we have%
\begin{align*}
  \short_{([m]\times [n]) \setminus (\cU\times \cV)}(\cC) \in \bbC_{\cU\times \cV}(a,b,0) \ .
\end{align*}
Further, for any $\cU\subseteq \cI$ for some $\cI \in \bbIcol$ and $\cV\subseteq \cI$ for some $\cI \in \bbIrow$ we have
\begin{align*}
\punct_{([m] \! \times \! [n]) \setminus (\cU \times  \cV)}(\cC)\! \in \!\bbC_{\cU \times \cV}(a\!-\!(m\!-\!u),b\!-\!(n\!-\!v),0) \ .
\end{align*}
\end{lemma}
\begin{IEEEproof}
  As $h=0$, by Definition~\ref{def:codesForGridLike}, the generator matrix of $\cC$ is given by $\Gcol \otimes \Grow$. Let $\Gcol^{\short_\cJ},\Grow^{\short_\cJ}$ and $\Gcol^{\punct_\cJ},\Grow^{\punct_\cJ}$ denote the generator matrices of the column/row code shortened/punctured in the positions indexed by $\cJ$. It follows directly from the definition of the Kronecker product that
  \begin{align*}
    \short_{([m]\times [n]) \setminus (\cU\times \cV)}(\cC) &= \myspan{\Gcol^{\short_{\cU}} \otimes \Grow^{\short_{\cV}}} \\
    \punct_{([m]\times [n]) \setminus (\cU\times \cV)}(\cC) &= \myspan{\Gcol^{\punct_{\cU}} \otimes \Grow^{\punct_{\cV}}}  \ .
  \end{align*}
  The statement follows from observing that shortening a position of an information set of an $[m,m-b]$ code gives an $[m-1,m-b-1]$ code. Similarly, puncturing a position in the complement of an information set gives an $[m-1,m-b]$ code.
\end{IEEEproof}

\begin{lemma}\label{lem:incorLargerTopology}
  If there exists a regular erasure pattern $\cE$ that is not correctable in $T_{m \times n}(a,b,0)$ then there exists a regular erasure pattern $\cE'$ that is not correctable in $T_{m+\delta \times n+\gamma}(a,b,0)$ for any $\delta,\gamma$.
\end{lemma}
\begin{IEEEproof}
  We show that $\cE'=\cE$ is not correctable\footnote{Even if $\cE$ is maximal for $T_{m \times n}(a,b,0)$, it is not maximal for the topology $T_{m+\delta \times n+\gamma}(a,b,0)$ for any $\delta,\gamma$. However, it is also easy to construct a non-correctable maximal pattern by having each additional column/row contain exactly $a$ or $b$ erasures, respectively.}. As $\cE'$ does not contain additional erasures compared to $\cE$ and the restriction of Definition~\ref{def:regularPattern} does not depend on $m$ or $n$, we conclude that the pattern is regular, i.e., the regularity of a pattern in $T_{m \times n}(a,b,0)$ directly implies the regularity of the same pattern in $T_{m+\delta \times n+\delta}(a,b,0)$ for any $\delta,\gamma$.

 By Lemma~\ref{lem:topologyRelation} we have $\short_{([m+\delta]\times [n+\gamma])\setminus ([m]\times [n])}(\cC') \in \bbC_{m\times n}(a,b,0)$ for any code $\cC' \in \bbC_{m+\delta\times n+\gamma}(a,b,0)$. By assumption, the restricted pattern $\cE' \cap ([m] \times [n]) = \cE$ is not correctable in $T_{m \times n}(a,b,0)$ and the statement follows from \ref{item:correctSubpattern}) in Proposition~\ref{prop:shortPunctProperties}.   %
\end{IEEEproof}

\begin{lemma}\label{lem:incorPuncturing}
  If there exists a regular erasure pattern $\cE$ that is not correctable in $T_{m \times n}(a,b,0)$ then there exists a regular erasure pattern $\cE'$ that is not correctable in $T_{m + \delta \times n + \gamma}(a+\delta,b+\gamma,0)$ for any $\delta,\gamma$.
\end{lemma}
\begin{IEEEproof}
  Denote by $\cCcol'$ and $\cCrow'$ the column and row code of a code $\cC' \in \bbC_{m + \delta \times n + \gamma}(a+\delta,b+\gamma,0)$. Without loss of generality assume that the first $m-a$ and $n-b$ positions are a (subset of an) information set of the code $\cCcol'$ and $\cCrow'$. Let $\cE'$ be the erasure pattern obtained by adding $\delta$ rows and $\gamma$ columns of erasures to $\cE$, i.e., $\cE' = \cE \cup \big(([m+\delta]\times [n+\gamma]) \setminus ([m]\times [n])\big)$.
We show that this pattern is regular for the topology $T_{m + \delta \times n + \gamma}(a+\delta,b+\gamma,0)$, i.e., fulfills Definition~\ref{def:regularPattern} for any $\cU'\subseteq [m+\delta], |\cU|=u\geq a+\delta$ and $\cV'\subseteq [n+\gamma], |\cV'|=v\geq b+\gamma$. As the $\delta/\gamma$ additional rows/columns consist only of erasures it suffices to show that the subsets of rows/columns with $[m+1,m+\delta] \subset \cU$ and $[n+1,n+\gamma] \subset \cV$ fulfill the condition. Define $\cU = \cU' \cap [m]$ and $\cV = \cV'\cap [n]$ and observe that $\cE'$ can be partitioned into two disjoint subsets with
  \begin{align*}
    |\cE' \cap (\cU \times \cV)| &\leq (v-\gamma) a + (u-\delta)b - ab \\
    | \cE' \cap ((\cU'\times \cV') \setminus (\cU \times \cV)) | &= v\delta + u\gamma- \delta\gamma \ ,
  \end{align*}
  where the first equality holds because $|\cE' \cap (\cU \times \cV)| = \cE$ is regular by definition.
  Then
  \begin{align*}
    |\cE' &\cap (\cU' \! \times \!\cV')| \!= \!|\cE' \cap (\cU\! \times\! \cV)|\! +\! | \cE' \cap ((\cU'\!\times\! \cV') \setminus (\cU \!\times\! \cV)) |  \\
&\leq ((v-\gamma) a + (u-\delta)b - ab) + (v\delta + u\gamma- \delta\gamma) \\ %
    &= v(a+\delta) + u(b+\gamma) - (a+\delta)(b+\gamma)
  \end{align*}
  and it follows that the pattern is regular for the topology $T_{m + \delta \times n + \gamma}(a+\delta,b+\gamma,0)$.

  By Lemma~\ref{lem:topologyRelation} we have $\punct_{([m+\delta]\times [n+\gamma]) \setminus ([m]\times [n])}(\cC') \in  \bbC_{m \times n}(a,b,0)$. By definition, the pattern $\cE =\cE' \setminus \big(([m+\delta]\times [n+\gamma]) \setminus ([m]\times [n])\big)$ is not correctable in this topology and the statement follows by \ref{item:incorPuncture}) in Proposition~\ref{prop:shortPunctProperties}.
\end{IEEEproof}

\begin{theorem}\label{thm:largerPatterns}
  Let $a,b\geq 2$. For any $m \geq a+3$ and $n\geq b+3$ there exist regular erasure patterns that are not correctable in the topology $T_{m \times n}(a,b,0)$.
\end{theorem}
\begin{IEEEproof}
  Follows from applying Lemma~\ref{lem:incorLargerTopology} and~\ref{lem:incorPuncturing} to the topology $T_{5 \times 5}(2,2,0)$, for which an uncorrectable regular erasure pattern exists by Lemma~\ref{lem:incorrectablePattern}.
\end{IEEEproof}

\section{Connection between Product and Tensor-Product Codes}

TP codes \cite{wolf1965codes} are duals of product codes and of interest for storage applications due to their small storage overhead and good protection against correlated errors.
However, there is no general classification of erasure patterns correctable by a TP code.
In this section, we show how the maximal erasure patterns of codes and their dual codes are connected.
This implies that for certain TP codes, which are duals of MR codes of a grid-like topology with $h=0$, we can exactly characterize their correctable erasure patterns.

\begin{definition}[Tensor-Product Code]
 Consider two codes $\cCcol[m,a]$ and $\cCrow[n,b]$. Then the tensor-product code $\mathsf{TP}(\cCcol,\cCrow)$ is the $[mn,mn-(m-a)(n-b)]$ code defined as
    $\mathsf{TP}(\cCcol,\cCrow) = \myspan{\Hcol \otimes \Hrow}^\perp$,
  where $\Hcol$ and $\Hrow$ denote parity-check matrices of the codes $\cCcol$ and $\cCrow$, respectively.
\end{definition}

\begin{lemma}
  Let $\cC_1$ be an $[m,a]$ code and $\cC_2$ be an $[n,b]$ code. Then $\mathsf{TP}(\cC_1,\cC_2)^\perp \in \bbC_{m \times n}(a,b,0)$.
\end{lemma}
\begin{IEEEproof}
  Denote by $\Hcol$ and $\Hrow$ parity-check matrices of the codes $\cCcol$ and $\cCrow$, respectively. We have
  $\sTP(\cCcol,\cCrow)^\perp = \myspan{\Hcol \otimes \Hrow}$ .
  The statement follows from interpreting the matrices $\Hcol$ and $\Hrow$ as generator matrices of $[m,m-a]$ and $[n,n-b]$ codes, respectively.
\end{IEEEproof}

\begin{lemma}
  Consider any code $\cC[n,k]$ and denote by $\bbE$ the set of erasure patterns that are correctable in this code. Then an erasure pattern $\cE$ is correctable in $\cC^\perp$ if and only if there exists a pattern $\cE_{\max}$ with $\cE\subseteq \cE_{\max}$ and $|\cE_{\max}| = n-k$ such that $[n]\setminus \cE_{\max} \in \bbE$.
\end{lemma}
\begin{IEEEproof}
  An erasure pattern is correctable if and only if it is the subset of the complement of an information set. Equivalently, an erasure pattern is correctable if and only if the corresponding positions are linearly independent in the dual code. This implies the lemma statement.
\end{IEEEproof}

\begin{corollary}
  The set of erasure patterns correctable in a code $\cC$ uniquely determines the set of erasure patterns correctable in the dual code $\cC^\perp$.
\end{corollary}

\begin{theorem}\label{thm:correctableTP}
  Let $\cCcol$ be an $[m,a]$ code and $\cCrow$ be an $[n,b]$ code. The set of maximal erasure patterns correctable by the TP code $\mathsf{TP}(\cCcol,\cCrow)$ is a subset of
  \begin{align*}
      \{([m] \times [n]) \setminus \cE \ | \ \cE \in  \bbE^{\max}_{m \times n}(a,b,0) \} \ .
  \end{align*}
  Moreover, if $\mathsf{TP}(\cCcol,\cCrow)^\perp \in  \bbCMR_{m \times n}(a,b,0)$ then it corrects all these erasure patterns.
\end{theorem}
Applying Theorem~\ref{thm:correctableTP} to the cases where $\bbE_{m \times n}(a,b,0)$ is known (see Table~\ref{tab:topologies}) establishes the correctable erasure patterns in the corresponding TP codes.

\bibliographystyle{IEEEtran}
\bibliography{main.bib}

\begin{thebibliography}{10}
\providecommand{\url}[1]{#1}
\csname url@samestyle\endcsname
\providecommand{\newblock}{\relax}
\providecommand{\bibinfo}[2]{#2}
\providecommand{\BIBentrySTDinterwordspacing}{\spaceskip=0pt\relax}
\providecommand{\BIBentryALTinterwordstretchfactor}{4}
\providecommand{\BIBentryALTinterwordspacing}{\spaceskip=\fontdimen2\font plus
\BIBentryALTinterwordstretchfactor\fontdimen3\font minus
  \fontdimen4\font\relax}
\providecommand{\BIBforeignlanguage}[2]{{%
\expandafter\ifx\csname l@#1\endcsname\relax
\typeout{** WARNING: IEEEtran.bst: No hyphenation pattern has been}%
\typeout{** loaded for the language `#1'. Using the pattern for}%
\typeout{** the default language instead.}%
\else
\language=\csname l@#1\endcsname
\fi
#2}}
\providecommand{\BIBdecl}{\relax}
\BIBdecl

\bibitem{gopalan2012locality}
P.~Gopalan, C.~Huang, H.~Simitci, and S.~Yekhanin, ``On the locality of
  codeword symbols,'' \emph{IEEE Transactions on Information theory}, vol.~58,
  no.~11, pp. 6925--6934, 2012.

\bibitem{rawat2013optimal}
A.~S. Rawat, O.~O. Koyluoglu, N.~Silberstein, and S.~Vishwanath, ``Optimal
  locally repairable and secure codes for distributed storage systems,''
  \emph{IEEE Transactions on Information Theory}, vol.~60, no.~1, pp. 212--236,
  2013.

\bibitem{chen2007maximally}
M.~Chen, C.~Huang, and J.~Li, ``On the maximally recoverable property for
  multi-protection group codes,'' in \emph{2007 IEEE International Symposium on
  Information Theory}.\hskip 1em plus 0.5em minus 0.4em\relax IEEE, 2007, pp.
  486--490.

\bibitem{blaum2016construction}
M.~Blaum, J.~S. Plank, M.~Schwartz, and E.~Yaakobi, ``Construction of partial
  {MDS} and sector-disk codes with two global parity symbols,'' \emph{IEEE
  Transactions on Information Theory}, vol.~62, no.~5, pp. 2673--2681, 2016.

\bibitem{blaum2013partial}
M.~Blaum, J.~L. Hafner, and S.~Hetzler, ``Partial-{MDS} codes and their
  application to {RAID} type of architectures,'' \emph{IEEE Transactions on
  Information Theory}, vol.~59, no.~7, pp. 4510--4519, 2013.

\bibitem{gopalan2014explicit}
P.~Gopalan, C.~Huang, B.~Jenkins, and S.~Yekhanin, ``Explicit maximally
  recoverable codes with locality,'' \emph{IEEE Transactions on Information
  Theory}, vol.~60, no.~9, pp. 5245--5256, 2014.

\bibitem{hu2016new}
G.~Hu and S.~Yekhanin, ``New constructions of {SD} and {MR} codes over small
  finite fields,'' in \emph{2016 IEEE International Symposium on Information
  Theory (ISIT)}.\hskip 1em plus 0.5em minus 0.4em\relax IEEE, 2016, pp.
  1591--1595.

\bibitem{gopalan2017maximally}
P.~Gopalan, G.~Hu, S.~Kopparty, S.~Saraf, C.~Wang, and S.~Yekhanin, ``Maximally
  recoverable codes for grid-like topologies,'' in \emph{Proceedings of the
  Twenty-Eighth Annual ACM-SIAM Symposium on Discrete Algorithms}.\hskip 1em
  plus 0.5em minus 0.4em\relax SIAM, 2017, pp. 2092--2108.

\bibitem{balaji2015partial}
S.~Balaji and P.~V. Kumar, ``On partial maximally-recoverable and
  maximally-recoverable codes,'' in \emph{2015 IEEE International Symposium on
  Information Theory (ISIT)}.\hskip 1em plus 0.5em minus 0.4em\relax IEEE,
  2015, pp. 1881--1885.

\bibitem{muralidhar2014f4}
S.~Muralidhar, W.~Lloyd, S.~Roy, C.~Hill, E.~Lin, W.~Liu, S.~Pan, S.~Shankar,
  V.~Sivakumar, L.~Tang \emph{et~al.}, ``f4: Facebook's warm $\{$BLOB$\}$
  storage system,'' in \emph{11th $\{$USENIX$\}$ Symposium on Operating Systems
  Design and Implementation ($\{$OSDI$\}$ 14)}, 2014, pp. 383--398.

\bibitem{shivakrishna2018maximally}
D.~Shivakrishna, V.~A. Rameshwar, V.~Lalitha, and B.~Sasidharan, ``On maximally
  recoverable codes for product topologies,'' in \emph{2018 Twenty Fourth
  National Conference on Communications (NCC)}.\hskip 1em plus 0.5em minus
  0.4em\relax IEEE, 2018, pp. 1--6.

\bibitem{kong2019new}
X.~Kong, J.~Ma, and G.~Ge, ``New bounds on the field size for maximally
  recoverable codes instantiating grid-like topologies,'' \emph{arXiv preprint
  arXiv:1901.06915}, 2019.

\bibitem{kane2019independence}
D.~Kane, S.~Lovett, and S.~Rao, ``The independence number of the {B}irkhoff
  polytope graph, and applications to maximally recoverable codes,'' \emph{SIAM
  Journal on Computing}, vol.~48, no.~4, pp. 1425--1435, 2019.

\bibitem{wolf1965codes}
J.~Wolf, ``On codes derivable from the tensor product of check matrices,''
  \emph{IEEE Transactions on Information Theory}, vol.~11, no.~2, pp. 281--284,
  1965.

\bibitem{calis2016general}
G.~Calis and O.~O. Koyluoglu, ``A general construction for {PMDS} codes,''
  \emph{IEEE Communications Letters}, vol.~21, no.~3, pp. 452--455, 2016.

\bibitem{berger2003isometries}
T.~P. Berger, ``Isometries for rank distance and permutation group of
  {Gabidulin} codes,'' \emph{IEEE Transactions on Information Theory}, vol.~49,
  no.~11, pp. 3016--3019, 2003.

\end{thebibliography}
\end{document}